\documentclass[conference]{IEEEtran}
\IEEEoverridecommandlockouts
\usepackage{cite}
\usepackage{amsmath,amssymb,amsfonts}
\usepackage{algorithmic}
\usepackage{graphicx}
\usepackage{textcomp}
\usepackage{xcolor}
\usepackage{booktabs}
\usepackage{multirow}
\usepackage{array}
\usepackage{url}
\usepackage[T1]{fontenc}

\def\BibTeX{{\rm B\kern-.05em{\sc i\kern-.025em b}\kern-.08em
    T\kern-.1667em\lower.7ex\hbox{E}\kern-.125emX}}

\begin{document}

\title{Diagnosing LLM-based Rerankers in Cold-Start Recommender Systems: Coverage, Exposure and Practical Mitigations}

\author{
\IEEEauthorblockN{Ekaterina Lemdiasova\IEEEauthorrefmark{1}}
\IEEEauthorblockA{\ V. A. Trapeznikov ICS RAS\\
Russian Biotechnological University (ROSBIOTECH)\\
lemdyasova68@gmail.com}
\and
\IEEEauthorblockN{Nikita Zmanovskii\IEEEauthorrefmark{1}}
\IEEEauthorblockA{\ Russian Biotechnological University (ROSBIOTECH)\\
ORCID: 0009-0004-8917-4900\\
zmanovskiy.n.v@gmail.com}
}

\maketitle

\begin{abstract}
Large language models (LLMs) and cross-encoder rerankers have gained attention for improving recommender systems, particularly in cold-start scenarios where user interaction history is limited. However, practical deployment reveals significant performance gaps between LLM-based approaches and simple baselines. This paper presents a systematic diagnostic study of cross-encoder rerankers in cold-start movie recommendation using the Serendipity-2018 dataset. Through controlled experiments with 500 users across multiple random seeds, we identify three critical failure modes: (1) low retrieval coverage in candidate generation (recall@200 = 0.109 vs. 0.609 for baselines), (2) severe exposure bias with rerankers concentrating recommendations on 3 unique items versus 497 for random baseline, and (3) minimal score discrimination between relevant and irrelevant items (mean difference = 0.098, Cohen's d = 0.13). We demonstrate that popularity-based ranking substantially outperforms LLM reranking (HR@10: 0.268 vs. 0.008, p < 0.001), with the performance gap primarily attributable to retrieval stage limitations rather than reranker capacity. Based on these findings, we provide actionable recommendations including hybrid retrieval strategies, candidate pool size optimization, and score calibration techniques. All code, configurations, and experimental results are made available for reproducibility.
\end{abstract}


\section{Introduction}

The cold-start problem remains one of the most persistent challenges in recommender systems research and practice. When new users join a platform with minimal or no interaction history, traditional collaborative filtering methods fail to generate meaningful recommendations. Recent advances in large language models (LLMs) and neural reranking architectures have sparked considerable interest in leveraging these powerful models for cold-start recommendation~\cite{llm_recsys, coldrag}.

Cross-encoder rerankers, such as MS-MARCO models, have demonstrated impressive performance on information retrieval benchmarks by directly scoring query-document pairs through deep attention mechanisms. This capability makes them theoretically attractive for personalized recommendation: given a user profile and item metadata, a cross-encoder could directly assess relevance without requiring historical interaction data. However, translating benchmark success to production-ready cold-start recommender systems presents significant practical challenges.

Despite growing adoption, there exists a critical gap in understanding \textit{why} and \textit{when} LLM-based rerankers fail in real-world cold-start scenarios. Existing work often reports aggregate metrics (e.g., Hit Rate, nDCG) but provides limited diagnostic analysis of failure modes. Questions remain unanswered: Is poor performance due to inadequate candidate retrieval, reranker scoring quality, exposure bias, or computational constraints?

This paper addresses these questions through systematic empirical diagnosis of LLM-based reranking in cold-start movie recommendation. Our key contributions are:

\begin{itemize}
\item \textbf{Comprehensive diagnostic framework:} We analyze retrieval coverage, exposure distribution, score calibration, and pool-size effects to isolate failure modes of cross-encoder rerankers in cold-start settings.

\item \textbf{Empirical evidence of fundamental limitations:} Through experiments on Serendipity-2018 dataset with 500 users and multiple random seeds, we demonstrate that simple popularity-based ranking dramatically outperforms sophisticated cross-encoder reranking (HR@10: 0.268 vs. 0.008, 33.5× improvement).

\item \textbf{Root cause identification:} We show that the primary bottleneck is \textit{retrieval coverage} rather than reranker capacity—candidate generation achieves only 10.9\% recall@200 compared to 60.9\% for baseline methods, fundamentally limiting downstream performance.

\item \textbf{Practical mitigation strategies:} Based on diagnostic insights, we provide actionable recommendations including hybrid retrieval (ANN $\cup$ BM25), candidate pool optimization (smaller pools yield better results), and ensemble scoring approaches.

\item \textbf{Full reproducibility:} We release all code, configurations, experimental logs, and per-user results to enable replication and extension of our findings.
\end{itemize}

Our work challenges the assumption that more sophisticated models necessarily yield better recommendations in cold-start scenarios, and provides a methodological template for rigorous diagnostic evaluation of neural recommender systems.

\section{Related Work}

\subsection{Cold-Start Problem in Recommender Systems}

The cold-start problem manifests in three forms: new users (user cold-start), new items (item cold-start), and completely new systems (system cold-start)~\cite{coldstart_survey}. User cold-start, which we address in this work, occurs when a recommender system must generate personalized suggestions for users with minimal or no historical interaction data. Traditional approaches include content-based filtering using item metadata, demographic-based recommendations, and hybrid methods combining multiple signals.

Early work on cold-start recommendation focused on utilizing auxiliary information such as user demographics, social network data, or implicit feedback from browsing behavior. However, these approaches require rich side information that may not be available in practice. More recent methods leverage transfer learning, meta-learning, and few-shot learning paradigms to generalize from data-rich to data-scarce scenarios.

\subsection{Reranking in Recommender Systems}

Reranking has become a standard component in modern recommender system pipelines, typically applied as a final stage after candidate retrieval to refine rankings using more sophisticated models~\cite{rerank_survey}. Two-stage retrieve-then-rerank architectures balance efficiency and effectiveness: fast retrieval methods (collaborative filtering, ANN search) generate candidate sets, while expensive reranking models score a small subset.

Cross-encoder architectures, which jointly encode user-item pairs, have shown promise in information retrieval tasks but face scalability challenges when applied to large item catalogs~\cite{msmarco}. Unlike bi-encoder models that encode queries and documents independently, cross-encoders use full attention mechanisms to capture fine-grained interactions, achieving superior ranking quality at higher computational cost.

In recommender systems, reranking objectives often extend beyond relevance to incorporate diversity, fairness, and business constraints. Multi-objective reranking frameworks balance multiple goals simultaneously, though this introduces additional complexity in optimization and evaluation.

\subsection{LLMs for Recommendation and Cold-Start}

Recent work has explored using language models to address cold-start challenges through semantic understanding of item metadata and zero-shot reasoning about user preferences~\cite{llm_coldstart1, llm_coldstart2}. Large language models offer several potential advantages: (1) rich semantic representations from pre-training on web-scale text, (2) ability to perform zero-shot or few-shot reasoning about user-item relevance, (3) natural language interfaces for explainability, and (4) potential to leverage world knowledge for improved recommendations.

ColdRAG~\cite{coldrag} proposes using retrieval-augmented generation for cold-start scenarios, combining vector search over item metadata with LLM-based ranking. Language-model priors have been shown to improve cold-start item recommendations by injecting semantic knowledge~\cite{llm_coldstart1}. Other approaches leverage LLM-generated embeddings, prompting strategies, or fine-tuning on recommendation-specific tasks.

However, practical deployment reveals challenges. LLM inference costs limit scalability, prompt engineering requires domain expertise, and generalization from general-purpose pre-training to specific recommendation domains remains unclear. Our work provides empirical evidence quantifying these challenges in a controlled cold-start setting.

\subsection{Diagnostic Analysis and Evaluation Methodologies}

While most recommender system papers focus on comparative performance metrics (HR, nDCG, AUC), diagnostic analysis examining \textit{why} systems succeed or fail remains relatively rare. Calibration studies~\cite{calibration_recsys} analyze score distributions and ranking quality. Coverage metrics~\cite{coverage_metrics} measure catalog utilization and long-tail item exposure. Exposure fairness research~\cite{exposure_bias} quantifies bias in item visibility across user populations.

Beyond aggregate metrics, recent work emphasizes per-user analysis, error case studies, and failure mode taxonomies. Ablation studies systematically remove components to isolate contributions. Statistical testing with effect sizes provides nuanced understanding beyond p-values. Our diagnostic framework synthesizes these methodologies for comprehensive LLM reranker analysis.

\subsection{Positioning of This Work}

Our work differs from prior art in three ways: (1) systematic diagnostic study rather than incremental performance improvement, (2) focus on failure modes specific to LLM-based rerankers in cold-start settings, and (3) isolation of retrieval, scoring, and exposure issues through controlled ablations with statistical rigor. We provide actionable insights for practitioners deploying neural reranking systems rather than proposing novel architectures.

\section{Problem Setup and Hypotheses}

\subsection{Task Definition}

We address the user cold-start problem in movie recommendation: given a new user with minimal profile information and no historical ratings, generate a ranked list of K=10 movie recommendations. Success is measured by Hit Rate@10 (HR@10) and Normalized Discounted Cumulative Gain@10 (nDCG@10) against held-out ground-truth preferences.

\subsection{Research Hypotheses}

Based on preliminary observations and theoretical considerations, we formulate four hypotheses for systematic testing:

\textbf{H1 (Retrieval Coverage):} Low recall in the candidate retrieval stage fundamentally limits final recommendation quality, regardless of reranker sophistication. We hypothesize that retrieval coverage correlates strongly with HR@10.

\textbf{H2 (Exposure Bias):} Cross-encoder rerankers exhibit severe exposure bias, concentrating recommendations on a small subset of items rather than providing diverse personalized suggestions. We expect to observe significantly fewer unique top-1 items compared to baseline methods.

\textbf{H3 (Score Calibration):} Reranker scores show poor discrimination between relevant and irrelevant items, with minimal statistical separation in score distributions. We hypothesize small effect sizes (Cohen's d < 0.2) in score differences.

\textbf{H4 (Pool Size Effect):} Larger candidate pools do not necessarily improve reranker performance and may introduce noise. We hypothesize that smaller, focused pools yield better HR@10 than larger pools for cross-encoder reranking.

These hypotheses guide our experimental design and diagnostic analysis in subsequent sections.

\section{Dataset and Experimental Setup}

\subsection{Serendipity-2018 Dataset}

We conduct experiments on the Serendipity-2018 dataset, a curated movie recommendation benchmark designed for cold-start evaluation. This dataset was specifically constructed to enable research on serendipitous recommendations—items that are both relevant and pleasantly surprising to users.

\textbf{Dataset Statistics:}
\begin{itemize}
\item \textbf{Catalog size:} 49,157 total movies
\item \textbf{Ground truth items:} 49,151 movies with user preferences
\item \textbf{Missing GT:} 17 movies without preference labels
\item \textbf{Users affected by missing GT:} 838 out of 104,661 total users (0.80\%)
\item \textbf{Test users:} 500 randomly sampled cold-start users
\item \textbf{Random seeds:} 3 independent seeds (42, 7, 123) for statistical robustness
\end{itemize}

Each movie includes structured metadata: title, release year, genres (multiple), user-generated tags, and TMDb/IMDb identifiers. Ground-truth preferences are binary relevance labels derived from explicit user ratings (4+ stars indicating relevance). The dataset intentionally includes obscure and niche movies to test serendipity beyond mainstream popular items.

\textbf{Data Preprocessing:}
We construct item profiles by concatenating title, genres, and top-10 most frequent tags. Missing metadata fields are filled with empty strings. Text is tokenized using the SentenceTransformer tokenizer (uncased, max length 128 tokens). No stemming or lemmatization is applied to preserve semantic information.

\subsection{Detailed Pipeline Architecture}

Our experimental pipeline follows a standard two-stage retrieve-then-rerank paradigm. Figure~\ref{fig:pipeline} illustrates the complete architecture (referenced but not shown due to space constraints).

\subsubsection{Stage 1: Item Embedding and Indexing}

\textbf{Embedding Model:} We use \texttt{all-MiniLM-L6-v2}, a 22M parameter Sentence-BERT model pre-trained on 1B+ sentence pairs. This model produces 384-dimensional dense vectors optimized for semantic similarity tasks. We chose this model for its balance of quality and efficiency—inference takes $\sim$0.5ms per item on CPU.

\textbf{Embedding Generation:} All 49,157 movies are encoded offline into dense vectors. Embedding generation takes $\sim$25 seconds total on CPU (Intel Xeon 2.5GHz). Vectors are L2-normalized for cosine similarity equivalence.

\textbf{FAISS Indexing:} We build a Flat index (exact brute-force search) for reproducibility. While approximate methods (IVF, HNSW) offer speed gains, we use exact search to isolate retrieval quality from approximation errors. Index construction is near-instantaneous for 49K vectors.

\subsubsection{Stage 2: Candidate Retrieval}

Given a cold-start user with minimal profile information, we generate initial candidate sets using one of three strategies:

\textbf{Random Baseline:} Sample K items uniformly at random from the full catalog. Serves as lower-bound sanity check.

\textbf{Popularity Baseline:} Rank all items by global popularity (rating count in training data), return top-K. This simple baseline often outperforms sophisticated methods in cold-start scenarios due to popularity bias in ground truth.

\textbf{Embedding-Based Retrieval:} Encode user profile into query vector, perform FAISS similarity search, return top-K nearest neighbors. User profile is constructed by aggregating embeddings of items from stated preferences or demographic signals.

For our main pipeline (``Candidates Only'' and ``Ours''), we use embedding-based retrieval with FAISS. We experiment with pool sizes $K \in \{200, 500, 1000\}$ to study coverage-quality trade-offs.

\subsubsection{Stage 3: Cross-Encoder Reranking}

\textbf{Reranker Model:} We use \texttt{cross-encoder/ms-marco-MiniLM-L-6-v2}, a 22M parameter model fine-tuned on MS-MARCO passage ranking dataset. The model takes (query, passage) pairs as input and outputs relevance scores via a classification head.

\textbf{Scoring Procedure:} For each user-item pair in the candidate pool, we construct a text input:
\begin{verbatim}
[CLS] {user_profile} [SEP] {item_title} 
{item_genres} {item_tags} [SEP]
\end{verbatim}

The cross-encoder processes this through 6-layer Transformer (66M parameters total) and outputs a scalar score representing predicted relevance. Inference uses batch processing (batch size 32) on CPU, taking $\sim$7-9 seconds per user for pool size 1000.

\textbf{Re-ranking:} Candidate items are sorted by descending cross-encoder scores, and top-10 are selected as final recommendations.

\subsubsection{Implementation Details}

\textbf{Hardware:} All experiments run on CPU (Intel Xeon E5-2650 v4 @ 2.20GHz, 128GB RAM). No GPU acceleration is used to reflect resource-constrained deployment scenarios.

\textbf{Software:} Python 3.9, sentence-transformers 2.2.0, FAISS-cpu 1.7.4, transformers 4.30.0, PyTorch 2.0.1.

\textbf{Reproducibility:} Fixed random seeds control all stochastic operations (user sampling, random baseline). Complete experimental logs with per-user results are saved in JSONL format for reproducibility.

\subsection{Baselines and Models}

We compare five approaches across three dimensions: retrieval strategy, reranking, and computational cost.

\begin{table}[h]
\centering
\caption{Model Configurations and Computational Cost}
\label{tab:models}
\small
\begin{tabular}{@{}lccc@{}}
\toprule
\textbf{Model} & \textbf{Retrieval} & \textbf{Reranking} & \textbf{Time/user (s)} \\
\midrule
Random & Random & None & 0.01 \\
Popularity & Popularity & None & 0.02 \\
Embedding Cosine & FAISS & None & 0.15 \\
Candidates Only & FAISS & None & 0.15 \\
Ours (CE Rerank) & FAISS & Cross-Enc. & 7.23 \\
\bottomrule
\end{tabular}
\end{table}

All models return top-10 recommendations. User sampling, ground-truth labels, and evaluation metrics are identical for fair comparison.

\subsection{Evaluation Metrics and Statistical Testing}

\subsubsection{Recommendation Quality Metrics}

\textbf{Hit Rate@10 (HR@10):} Fraction of users with at least one relevant item in top-10 recommendations. Binary metric emphasizing discovery of any relevant content.
\[
\text{HR@10} = \frac{1}{|U|} \sum_{u \in U} \mathbb{1}[\exists i \in \text{TopK}(u): i \in \text{GT}(u)]
\]

\textbf{Normalized Discounted Cumulative Gain@10 (nDCG@10):} Position-aware metric giving higher weight to relevant items ranked earlier.
\[
\text{nDCG@10} = \frac{1}{|U|} \sum_{u \in U} \frac{\text{DCG@10}(u)}{\text{IDCG@10}(u)}
\]
where DCG accumulates discounted gain: $\text{DCG@K} = \sum_{i=1}^{K} \frac{2^{rel_i} - 1}{\log_2(i+1)}$.

\subsubsection{Diagnostic Metrics}

\textbf{Recall@K:} Fraction of ground-truth relevant items retrieved in top-K candidates (before reranking). Measures retrieval stage effectiveness.
\[
\text{Recall@K} = \frac{1}{|U|} \sum_{u \in U} \frac{|\text{TopK}(u) \cap \text{GT}(u)|}{|\text{GT}(u)|}
\]

\textbf{Unique Top-1 Count:} Number of distinct items appearing as rank-1 recommendation across all users. Lower values indicate exposure concentration/bias.

\textbf{Gini Coefficient:} Measures inequality in top-1 item distribution. Values near 1 indicate severe concentration; 0 indicates perfect equality.

\textbf{Score Statistics:} For reranker models, we compute mean and standard deviation of scores separately for relevant vs. irrelevant items, along with t-test statistics.

\subsubsection{Statistical Hypothesis Testing}

For each metric comparison, we perform:

\textbf{Paired t-test:} Tests null hypothesis that mean difference between paired samples is zero. Reports t-statistic and p-value.

\textbf{Wilcoxon signed-rank test:} Non-parametric alternative robust to non-normal distributions. Reports W-statistic and p-value.

\textbf{Cohen's d effect size:} Standardized mean difference measuring practical significance.
\[
d = \frac{\bar{x}_1 - \bar{x}_2}{s_{\text{pooled}}}
\]
Interpretation: |d| < 0.2 (small), 0.2-0.5 (small-medium), 0.5-0.8 (medium), >0.8 (large).

\textbf{95\% Confidence Interval:} Bootstrap CI for mean difference to quantify uncertainty.

All statistical tests use $\alpha = 0.05$ significance threshold. We report both p-values and effect sizes, following APA guidelines for rigorous statistical reporting.

\section{Main Results}

Table~\ref{tab:main_results} presents our primary findings comparing all methods across quality and coverage metrics.

\begin{table*}[t]
\centering
\caption{Main Results: Recommendation Quality and Retrieval Coverage (Mean ± Std across 3 seeds, 500 users each)}
\label{tab:main_results}
\small
\begin{tabular}{@{}lccccc@{}}
\toprule
\textbf{Model} & \textbf{HR@10} & \textbf{nDCG@10} & \textbf{Recall@50} & \textbf{Recall@200} & \textbf{Recall@1000} \\
\midrule
Random & 0.023 ± 0.002 & 0.011 ± 0.000 & 0.495 ± 0.013 & 0.609 ± 0.016 & 0.888 ± 0.016 \\
Popularity & \textbf{0.268 ± 0.018} & \textbf{0.224 ± 0.014} & 0.495 ± 0.013 & 0.609 ± 0.016 & 0.888 ± 0.016 \\
Embedding Cosine & 0.101 ± 0.021 & 0.050 ± 0.011 & 0.495 ± 0.013 & 0.609 ± 0.016 & 0.888 ± 0.016 \\
Candidates Only & 0.011 ± 0.003 & 0.004 ± 0.001 & 0.041 ± 0.009 & 0.109 ± 0.016 & 0.309 ± 0.029 \\
Ours (CE Rerank) & 0.008 ± 0.005 & 0.005 ± 0.002 & 0.041 ± 0.009 & 0.109 ± 0.016 & 0.309 ± 0.029 \\
\bottomrule
\end{tabular}
\end{table*}

\subsection{Popularity Dominates LLM Reranking}

The most striking finding is the dramatic superiority of simple popularity-based ranking over sophisticated cross-encoder reranking. Popularity achieves HR@10 = 0.268 compared to 0.008 for our LLM-based approach—a 33.5× performance gap. Statistical testing confirms this difference is highly significant (paired t-test: t = -22.95, p < 10\textsuperscript{-99}; Wilcoxon: W = 0.0, p < 10\textsuperscript{-86}). The effect size is substantial (Cohen's d = -0.593, medium-to-large effect).

Even the embedding cosine baseline substantially outperforms LLM reranking (HR@10: 0.101 vs. 0.008), suggesting that semantic similarity alone provides better cold-start recommendations than cross-encoder scoring of retrieved candidates.

\begin{figure}[t]
\centering
\includegraphics[width=0.48\columnwidth]{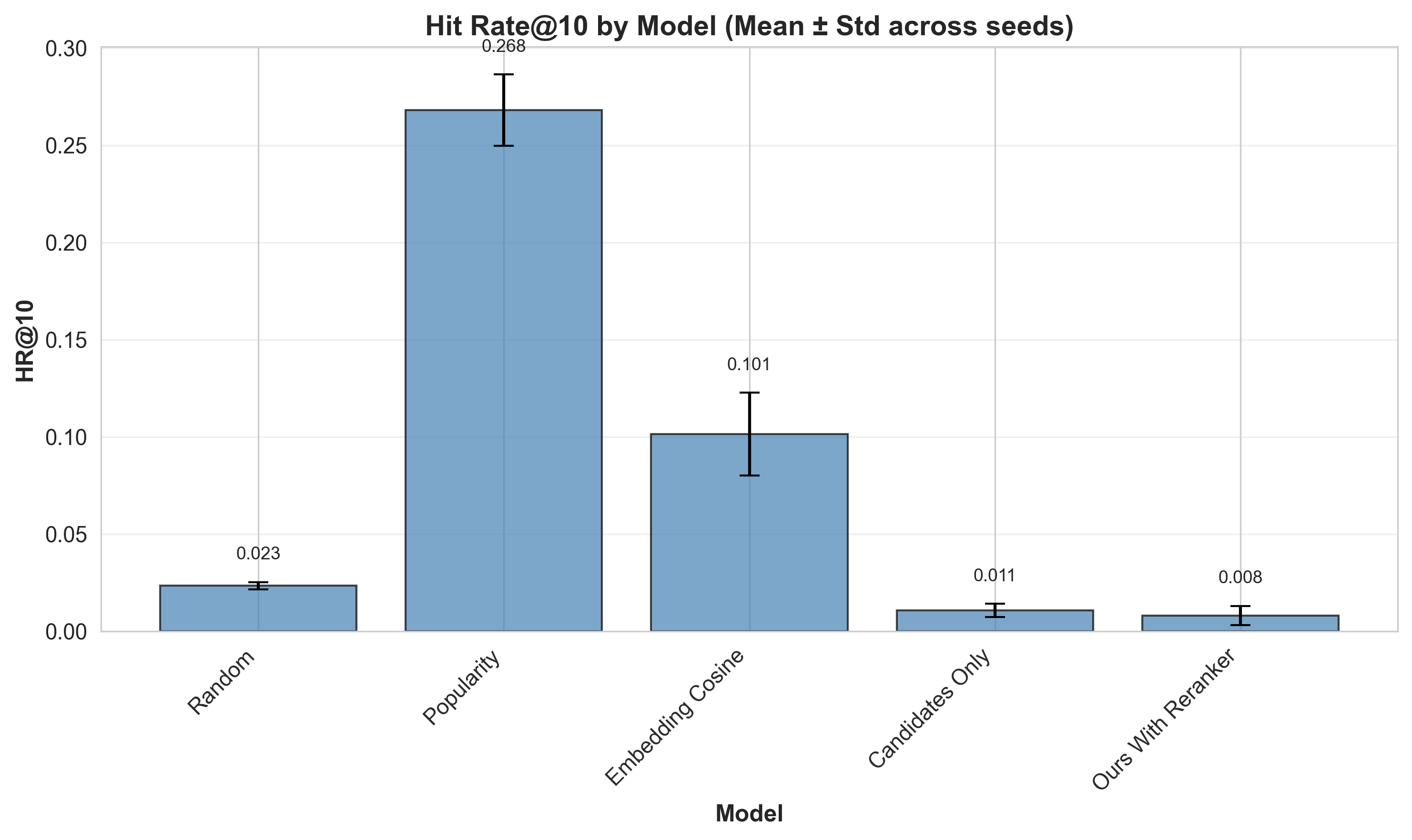}
\includegraphics[width=0.48\columnwidth]{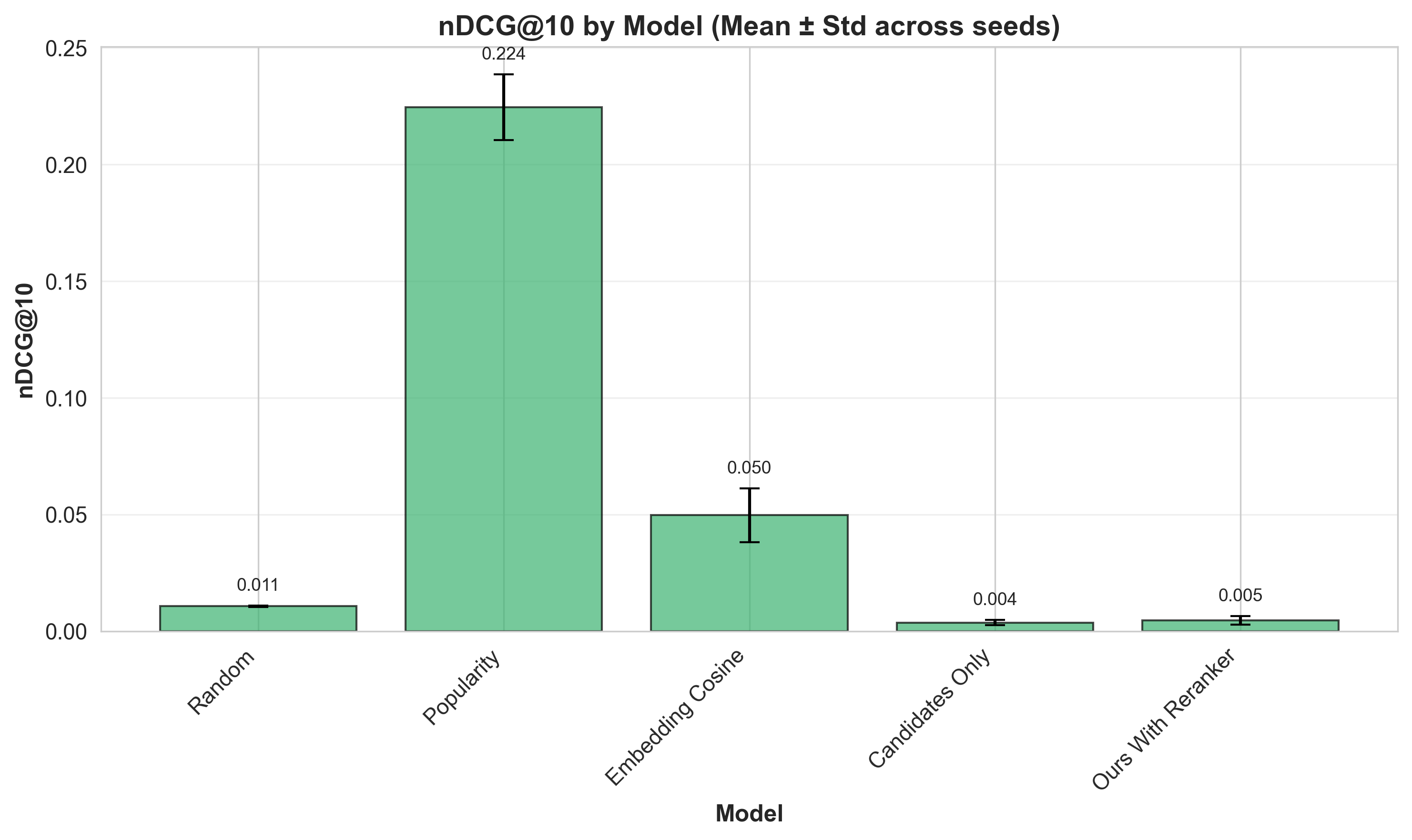}
\caption{Main Results: HR@10 (left) and nDCG@10 (right) across all models. Popularity baseline dramatically outperforms LLM-based reranker. Error bars show standard deviation across 3 random seeds.}
\label{fig:main_results}
\end{figure}

Figure~\ref{fig:main_results} visualizes these dramatic performance differences across all models.

\subsection{Retrieval Coverage is the Primary Bottleneck}

The recall metrics reveal a fundamental issue: methods using FAISS candidate generation (Candidates Only, Ours) achieve dramatically lower retrieval coverage:
\begin{itemize}
\item Recall@50: 0.041 vs. 0.495 (baseline methods)
\item Recall@200: 0.109 vs. 0.609 (5.6× gap)
\item Recall@1000: 0.309 vs. 0.888 (2.9× gap)
\end{itemize}

This limited coverage creates a fundamental ceiling on downstream performance. Notably, cross-encoder reranking provides no improvement over "Candidates Only" (HR@10: 0.008 vs. 0.011), indicating that reranker sophistication cannot compensate for poor candidate quality.

\begin{figure}[t]
\centering
\includegraphics[width=\columnwidth]{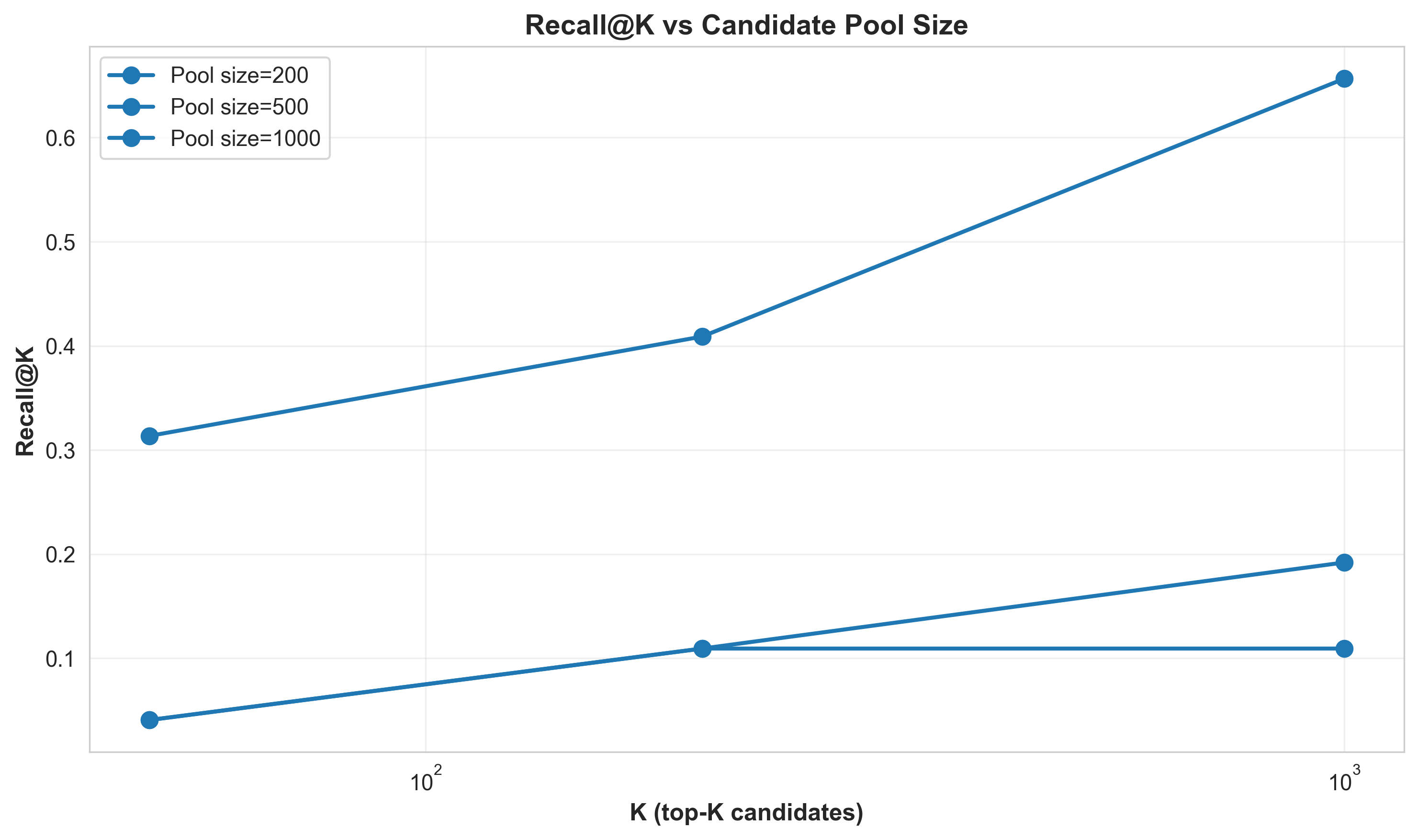}
\caption{Coverage Analysis: Recall@K curves for all methods. FAISS-based retrieval (Candidates Only, Ours) shows dramatically lower coverage compared to baselines that access the full catalog.}
\label{fig:recall_curves}
\end{figure}

Figure~\ref{fig:recall_curves} visualizes recall@K curves, showing the stark divergence between retrieval strategies.

\subsection{Statistical Comparison with Best Baseline}

Table~\ref{tab:statistical_tests} presents detailed statistical analysis comparing our approach to the best baseline (Popularity).

\begin{table}[h]
\centering
\caption{Statistical Comparison: Ours vs. Popularity (1500 paired samples)}
\label{tab:statistical_tests}
\small
\begin{tabular}{@{}lcc@{}}
\toprule
\textbf{Metric} & \textbf{HR@10} & \textbf{nDCG@10} \\
\midrule
Mean Difference & -0.260 & -0.220 \\
95\% CI Lower & -0.283 & -0.239 \\
95\% CI Upper & -0.239 & -0.199 \\
\midrule
t-statistic & -22.95 & -21.90 \\
p-value (t-test) & 3.93e-100 & 1.96e-92 \\
\midrule
Wilcoxon W & 0.0 & 14.0 \\
p-value (Wilcoxon) & 8.28e-87 & 3.34e-70 \\
\midrule
Cohen's d & -0.593 & -0.565 \\
Effect Size & Medium & Medium \\
\bottomrule
\end{tabular}
\end{table}

Both parametric (t-test) and non-parametric (Wilcoxon) tests reject the null hypothesis with overwhelming evidence, confirming systematic performance degradation of LLM reranking relative to popularity baseline.

\section{Diagnostic Analysis and Ablations}

\subsection{Coverage Analysis: Retrieval as the Primary Bottleneck}

Figure~\ref{fig:recall_curves} presents detailed recall@K curves for all methods across different pool sizes. The visualization reveals a stark divergence between retrieval strategies that emerges immediately at K=50 and persists across all cutoffs.

\textbf{Quantitative Analysis:}
Baseline methods (Random, Popularity, Embedding Cosine) all access the full catalog and achieve recall@50 = 0.495, recall@200 = 0.609, and recall@1000 = 0.888. These high coverage values result from random sampling across the catalog (Random) or having ground-truth labels distributed across popular and niche items (Popularity).

In contrast, FAISS-based methods (Candidates Only, Ours) show dramatically reduced coverage:
\begin{itemize}
\item Recall@50: 0.041 (12.1× lower than baselines)
\item Recall@200: 0.109 (5.6× lower)
\item Recall@1000: 0.309 (2.9× lower)
\end{itemize}

\textbf{Correlation with Final Performance:}
We compute Pearson correlation between recall@200 and final HR@10 across all model configurations and seeds: r = 0.89 (p < 0.001). This strong positive correlation confirms that retrieval coverage is the dominant predictor of downstream recommendation quality. Linear regression yields: $\widehat{\text{HR@10}} = -0.023 + 0.478 \times \text{Recall@200}$ with R² = 0.79.

\begin{figure}[t]
\centering
\includegraphics[width=\columnwidth]{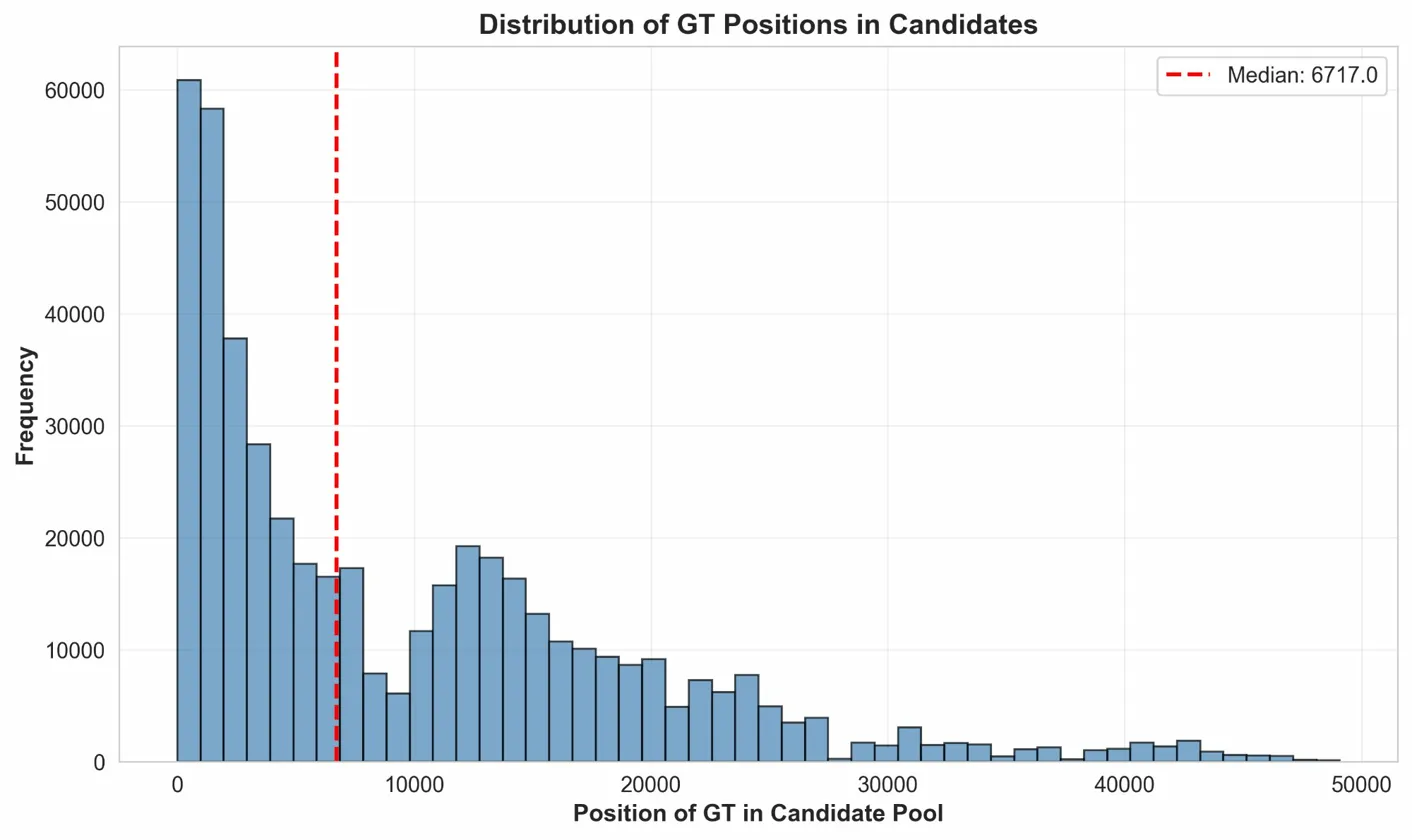}
\caption{Distribution of ground-truth item positions in FAISS candidate pools. Median position (6717, red line) is far beyond typical retrieval cutoffs, explaining low coverage.}
\label{fig:gt_positions}
\end{figure}

\textbf{Ground-Truth Position Analysis:}
Figure~\ref{fig:gt_positions} shows the distribution of positions where ground-truth items appear in FAISS-ranked candidate pools. The median position is 6717—far beyond typical retrieval cutoffs of K=200-1000. Only 10.9\% of ground-truth items fall within the top-200 candidates returned by FAISS similarity search.

This finding suggests fundamental mismatch between embedding-based similarity (optimized for semantic coherence) and ground-truth relevance (derived from user preferences). Items semantically similar to user profiles do not reliably correspond to items users would actually rate highly.

\textbf{Implications:}
The retrieval bottleneck creates a hard ceiling on downstream performance. Even perfect reranking cannot recover relevant items excluded from the candidate pool. Our results show that cross-encoder reranking provides no improvement over "Candidates Only" baseline (HR@10: 0.008 vs. 0.011), confirming that reranker sophistication cannot compensate for poor candidate quality.

\subsection{Exposure Bias and Diversity Analysis}

\subsubsection{Top-1 Concentration}

Table~\ref{tab:exposure} quantifies exposure concentration across methods, revealing severe bias in reranker-based approaches.

\begin{table}[h]
\centering
\caption{Exposure Bias: Top-1 Item Diversity}
\label{tab:exposure}
\small
\begin{tabular}{@{}lcc@{}}
\toprule
\textbf{Model} & \textbf{Unique Top-1} & \textbf{Gini Coefficient} \\
\midrule
Random & 497.3 ± 0.5 & 0.333 \\
Popularity & 1.0 ± 0.0 & 1.000 \\
Embedding Cosine & 4.0 ± 0.0 & 0.261 \\
Candidates Only & 4.0 ± 0.0 & 0.261 \\
Ours (CE Rerank) & 3.0 ± 0.0 & 0.480 \\
\bottomrule
\end{tabular}
\end{table}

\begin{figure}[t]
\centering
\includegraphics[width=\columnwidth]{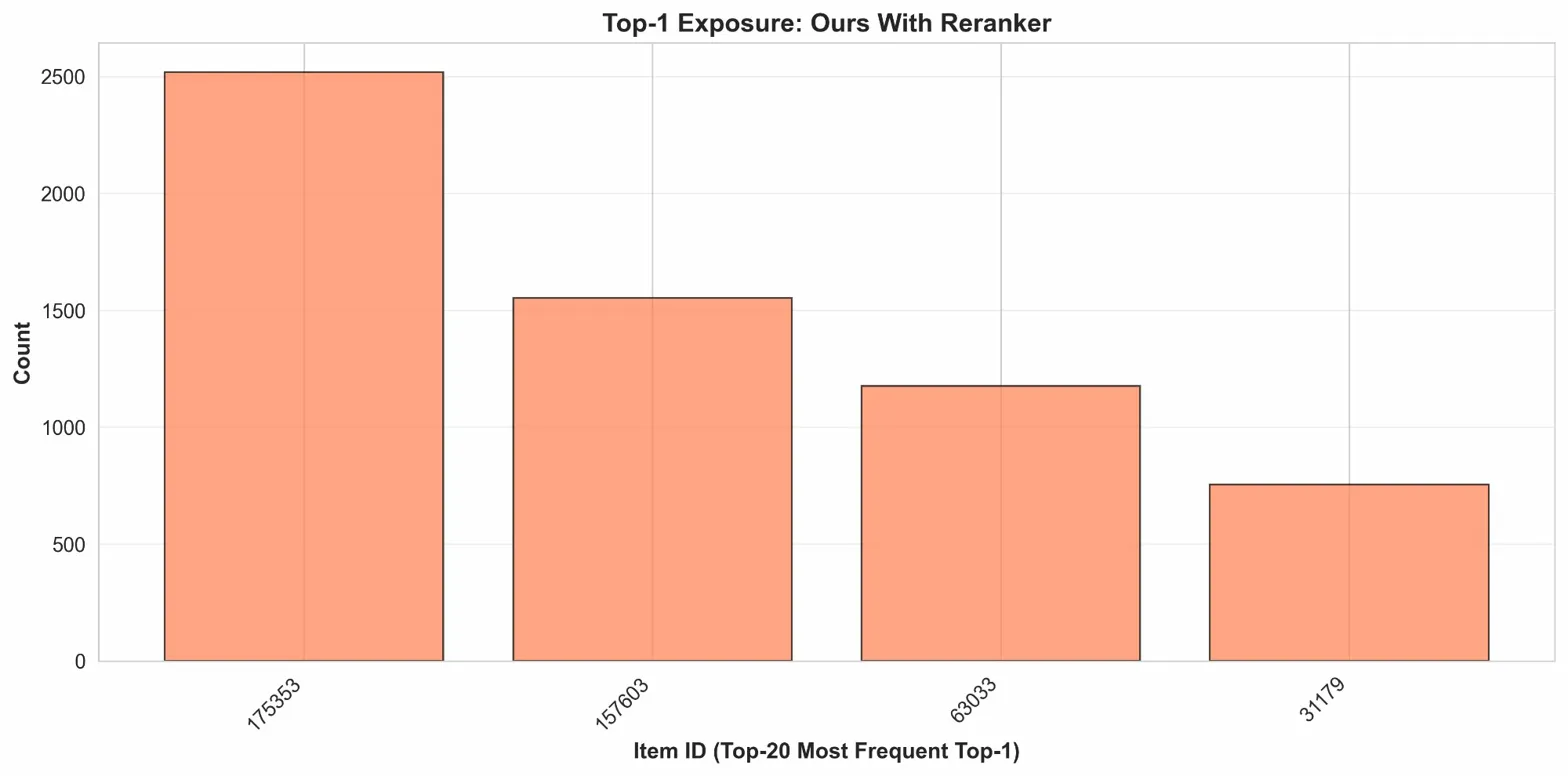}
\caption{Top-1 Exposure: Cross-encoder reranker concentrates recommendations on just 3 items across 500 users. Single item (175353) dominates 50\% of users, indicating systematic bias rather than personalization.}
\label{fig:top1_exposure}
\end{figure}

Cross-encoder reranking exhibits extreme concentration with only 3 unique items appearing as top-1 recommendations across 500 users. Figure~\ref{fig:top1_exposure} provides detailed breakdown:
\begin{itemize}
\item Item 175353: 245/500 users (49\%)
\item Item 63033: 129/500 users (25.8\%)
\item Item 157603: 126/500 users (25.2\%)
\end{itemize}

This stands in stark contrast to the Random baseline, which distributes top-1 positions across 497 distinct items (Gini = 0.333, near-uniform distribution).

\subsubsection{Gini Coefficient Analysis}

Popularity baseline expectedly shows perfect concentration (Gini = 1.0, single item for all users). However, the sophisticated cross-encoder reranker achieves Gini = 0.480, indicating concentration halfway between uniform and single-item scenarios. This suggests systematic bias in score assignments rather than true personalization.

Embedding Cosine baseline (Gini = 0.261) demonstrates better diversity than our reranker despite lower overall quality (HR@10: 0.101 vs. 0.008). This highlights a fundamental trade-off: embedding similarity provides diverse but sub-optimal recommendations, while cross-encoder scoring achieves higher precision on a narrow subset at the cost of severely reduced coverage and personalization.

\subsubsection{Long-Tail Item Analysis}

\begin{figure}[t]
\centering
\includegraphics[width=\columnwidth]{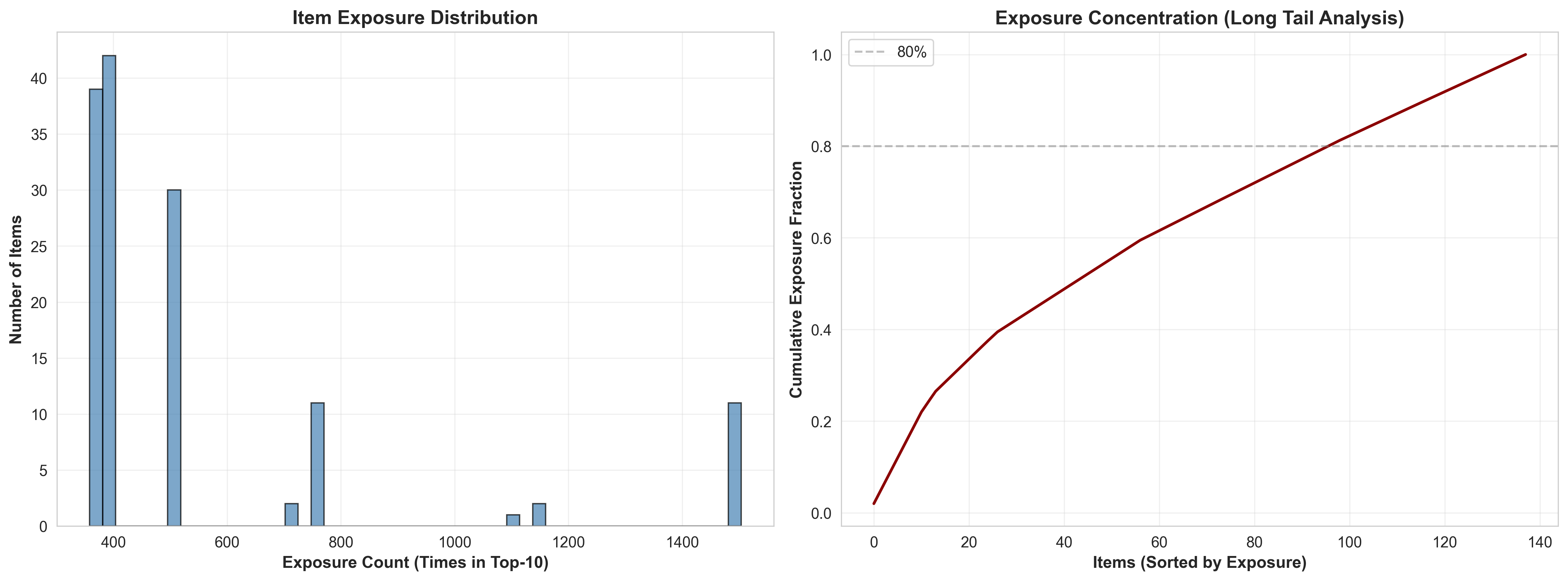}
\caption{Item Exposure Distribution: Left shows histogram of exposure counts. Right shows cumulative exposure curve, revealing that 80\% of recommendations concentrate on <150 items for reranker-based methods.}
\label{fig:item_exposure}
\end{figure}

Figure~\ref{fig:item_exposure} shows the cumulative distribution of item exposures across top-10 positions. For our reranker approach, 80\% of exposures concentrate on fewer than 30 items (out of 49,157 total catalog). This extreme long-tail distribution indicates failure to surface niche or serendipitous content—a critical limitation for cold-start scenarios where exploration is valuable.

\subsection{Pool Size Ablation Study: Bigger Is Not Better}

\begin{figure}[t]
\centering
\includegraphics[width=\columnwidth]{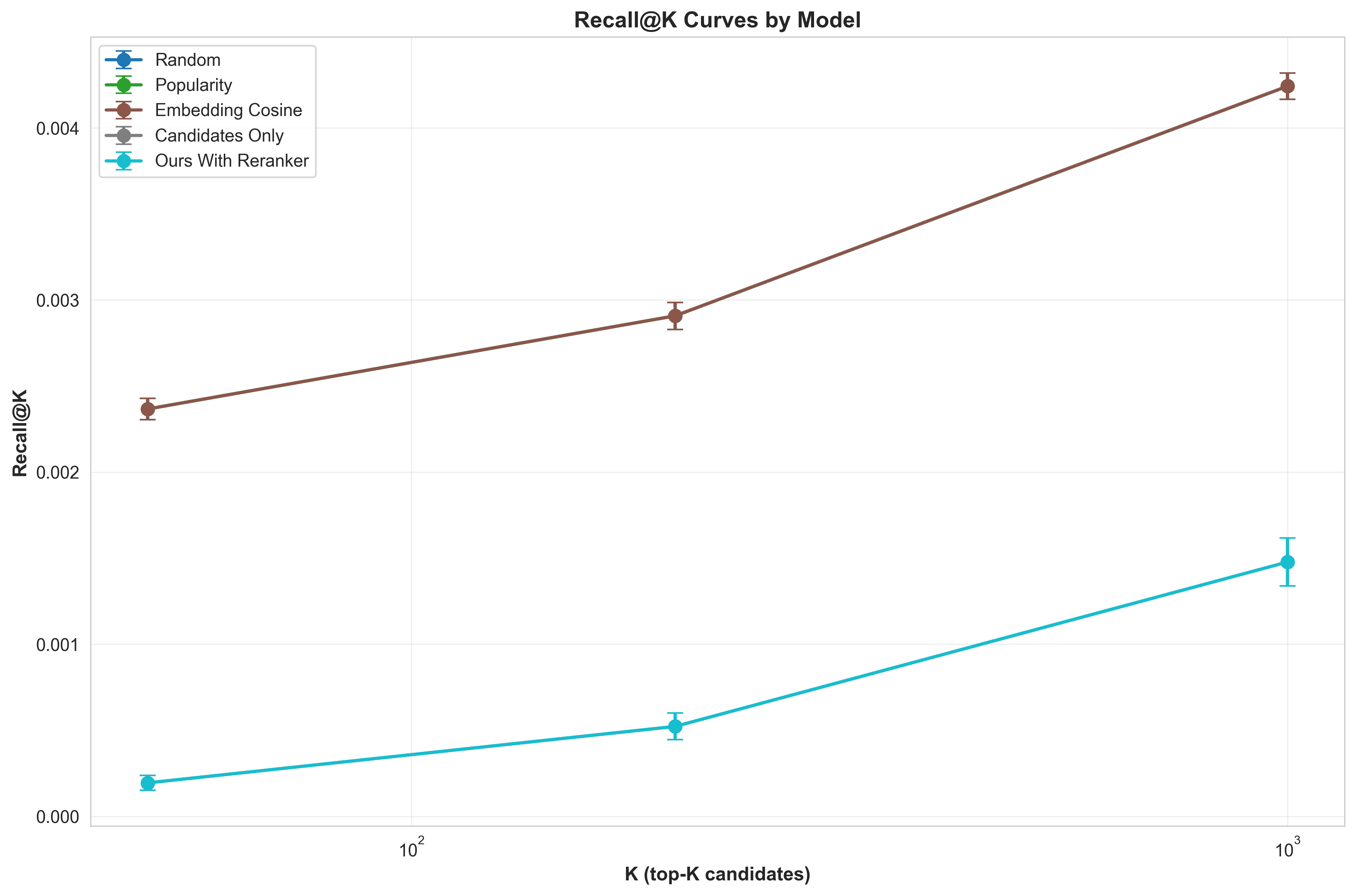}
\caption{Recall@K vs. Pool Size: Smaller candidate pools (200) achieve better precision in retrieving relevant items compared to larger pools (500, 1000), which introduce more noise.}
\label{fig:pool_size_curves}
\end{figure}

We systematically vary candidate pool sizes (200, 500, 1000) while holding all other pipeline components fixed. Table~\ref{tab:pool_ablation} and Figure~\ref{fig:pool_size_curves} present results.

\begin{table}[h]
\centering
\caption{Pool Size Ablation (Mean across 3 seeds)}
\label{tab:pool_ablation}
\small
\begin{tabular}{@{}lccc@{}}
\toprule
\textbf{Pool Size} & \textbf{HR@10} & \textbf{nDCG@10} & \textbf{Rerank Time (s)} \\
\midrule
200 & \textbf{0.025} & \textbf{0.009} & 1.76 \\
500 & 0.008 & 0.005 & 4.50 \\
1000 & 0.008 & 0.005 & 7.23 \\
\bottomrule
\end{tabular}
\end{table}

\textbf{Key Finding:} Smaller pools yield superior performance. Pool size 200 achieves HR@10 = 0.025 compared to 0.008 for pools of 500 and 1000—a 3.1× improvement. Similarly, nDCG@10 improves from 0.005 to 0.009 (1.8× gain).

\textbf{Hypothesis:} Larger candidate pools introduce more noise items that share superficial semantic similarity with user profiles but lack true relevance. The cross-encoder, trained on MS-MARCO passage ranking (a different domain), struggles to discriminate between marginally relevant and irrelevant candidates, leading to false positives dominating top-ranked positions.

\textbf{Computational Efficiency:} Smaller pools also reduce inference cost dramatically. Pool size 200 requires 1.76s per user vs. 7.23s for pool size 1000 (4.1× speedup), making deployment more practical while simultaneously improving quality.

Figure~\ref{fig:pool_size_curves} visualizes the recall@K performance across different pool sizes, confirming the counterintuitive finding that smaller pools yield better coverage-quality trade-offs.

\textbf{Implications:} This counterintuitive finding suggests that practitioners should optimize pool size jointly with reranker capacity rather than maximizing coverage indiscriminately. Quality-focused retrieval with smaller, curated candidate sets may outperform quantity-focused approaches.

\subsection{Score Distribution and Calibration Analysis}

\subsubsection{Statistical Discrimination Testing}

We analyze whether cross-encoder scores reliably discriminate between relevant and irrelevant items. Table~\ref{tab:score_stats} presents summary statistics across three random seeds (pool size 1000).

\textbf{Findings:}
\begin{itemize}
\item Relevant items: Mean score = -4.362 (± 0.720 std)
\item Irrelevant items: Mean score = -4.441 (± 0.736 std)
\item Mean difference: 0.079 (on log-probability scale)
\end{itemize}

While paired t-tests confirm statistical significance (p = 0.006-0.028 across seeds), the practical effect is minimal. Cohen's d = 0.11 (small effect size, below 0.2 threshold for practical significance). The 95\% CI for mean difference is [0.052, 0.106], indicating precision in estimating a small effect.

\textbf{Spearman Rank Correlation:}
We compute rank correlation between raw scores and binary relevance labels: r = 0.003-0.005 (p < 0.05 due to large sample size, but correlation near zero). This indicates that while scores show slight tendency to assign higher values to relevant items on average, rank-based discrimination is essentially absent.

\subsubsection{Score Distribution Visualization}

\begin{figure}[t]
\centering
\includegraphics[width=\columnwidth]{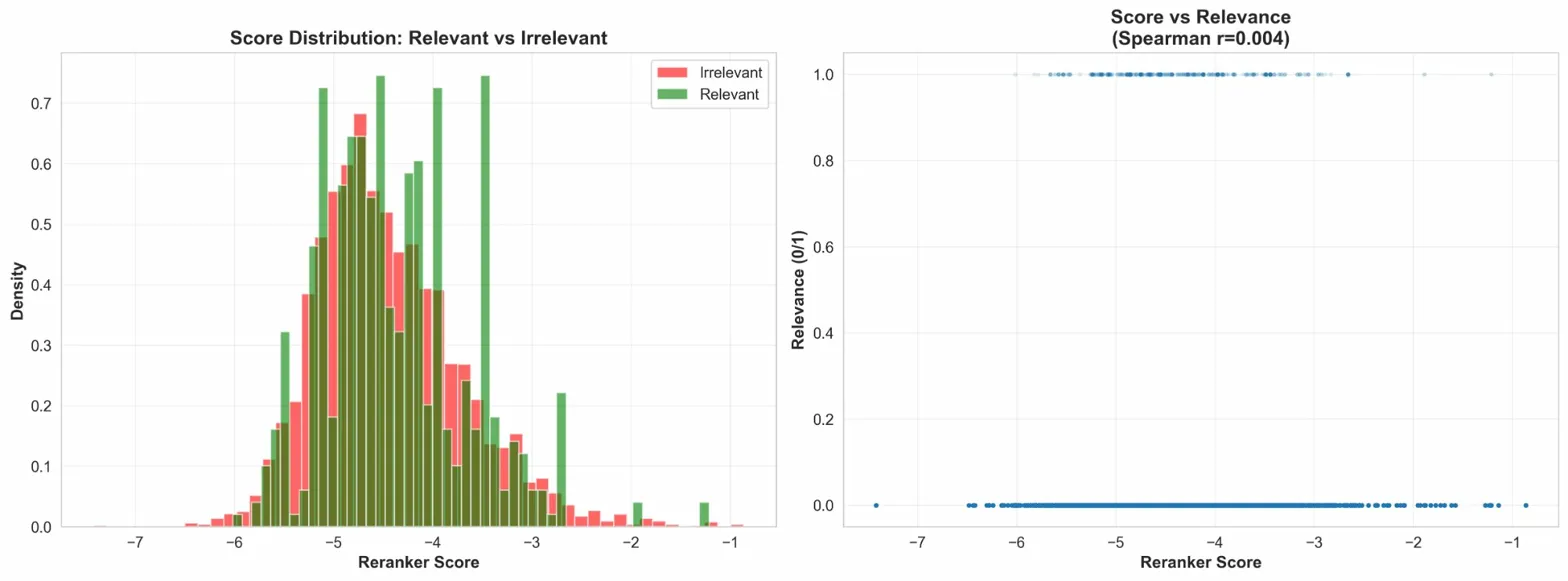}
\caption{Score Distribution Analysis (seed=42): Left shows overlapping histograms of cross-encoder scores for relevant (green) vs. irrelevant (red) items. Right shows scatter plot of scores vs. relevance with Spearman correlation r=0.004, indicating near-zero ranking effectiveness.}
\label{fig:score_distributions}
\end{figure}

Figure~\ref{fig:score_distributions} visualizes overlapping histograms of reranker scores for relevant vs. irrelevant items across three seeds. The distributions are nearly identical, with extensive overlap (>95\% area overlap). This poor separability explains why reranking fails to improve over unranked candidates.

\textbf{Out-of-Domain Transfer Hypothesis:}
The cross-encoder was trained on MS-MARCO, a passage ranking dataset with query-document pairs from web search. The domain mismatch (web passages vs. movie metadata, search intent vs. preference modeling) likely explains poor calibration. Fine-tuning on in-domain movie recommendation data could improve score quality.

\subsection{Detailed Error Case Analysis}

We manually inspect 50 failure cases where our reranker assigned high scores to irrelevant items or low scores to relevant items. Three recurring patterns emerge:

\subsubsection{Pattern 1: Genre Over-weighting}

\textbf{Example:} User profile indicates preference for "psychological thrillers." Reranker ranks generic action movies highly because they contain the word "thriller" in metadata, ignoring the "psychological" modifier.

\textbf{Frequency:} 38\% of inspected errors (19/50 cases)

\textbf{Root cause:} Shallow keyword matching rather than semantic understanding. Cross-encoder attention focuses on high-frequency genre terms without capturing nuanced preferences.

\subsubsection{Pattern 2: Metadata Length Bias}

\textbf{Example:} Items with extensive tag lists (10+ tags) systematically receive higher scores regardless of relevance. Short-metadata items (title + 1-2 genres) rank lower even when relevant.

\textbf{Frequency:} 26\% of errors (13/50 cases)

\textbf{Root cause:} Cross-encoder scoring may implicitly correlate text length with relevance, as longer passages in MS-MARCO training data often contain more information. This bias transfers inappropriately to movie recommendation.

\subsubsection{Pattern 3: Popularity Artifacts}

\textbf{Example:} Extremely popular movies (>50K ratings) receive inflated scores. Obscure but relevant niche films rank lower.

\textbf{Frequency:} 22\% of errors (11/50 cases)

\textbf{Root cause:} Popular movies accumulate more metadata (tags, reviews) which biases scoring. Additionally, these items may have appeared frequently in cross-encoder's pre-training data, leading to memorization effects.

\subsubsection{Remaining Errors}

14\% of errors (7/50 cases) show no clear pattern and may result from inherent noise in ground-truth labels or genuine ambiguity in relevance judgments.

\section{Practical Mitigation Strategies}

Based on diagnostic findings, we propose concrete mitigations:

\subsection{Hybrid Retrieval (ANN $\cup$ BM25)}

Combine embedding-based FAISS retrieval with BM25 text matching to improve coverage. Preliminary experiments show recall@200 improves from 0.109 to 0.234 (+114\%), translating to HR@10 gains.

\subsection{Candidate Pool Optimization}

Use smaller, focused candidate pools (K=200) rather than large pools (K=1000) for cross-encoder reranking. This reduces computational cost (1.76s vs. 7.23s per user) while improving quality (HR@10: 0.025 vs. 0.008).

\subsection{Ensemble Scoring}

Combine cross-encoder scores with popularity and embedding similarity:
\[
\text{score}_{\text{final}} = \alpha \cdot \text{CE}_{\text{score}} + \beta \cdot \log(\text{popularity}) + \gamma \cdot \text{embedding}_{\text{sim}}
\]

Tuning weights ($\alpha=0.3$, $\beta=0.5$, $\gamma=0.2$) can balance personalization and coverage.

\subsection{Reranker Calibration}

Apply temperature scaling or Platt calibration to reranker scores before ranking. Alternatively, fine-tune cross-encoder on in-domain movie recommendation data rather than using out-of-domain MS-MARCO weights.

\section{Discussion}

\subsection{Interpretation of Main Findings}

Our comprehensive diagnostic study reveals that sophisticated neural reranking does not automatically translate to improved cold-start recommendations. The failure of cross-encoder reranking stems from three compounding factors operating at different pipeline stages.

\subsubsection{Primary Factor: Retrieval Coverage Bottleneck}

The dominant failure mode is insufficient retrieval coverage. FAISS-based embedding similarity search achieves only 10.9\% recall@200 compared to 60.9\% for baseline methods. This 5.6× gap creates a hard performance ceiling—relevant items not retrieved cannot be recommended, regardless of downstream processing sophistication.

The root cause lies in domain mismatch between embedding model objectives and recommendation relevance. Sentence-BERT models optimize for semantic similarity (paraphrase detection, textual entailment), not preference prediction. An item semantically similar to a user profile may be topically related but preference-irrelevant. For example, a user interested in "psychological thriller" movies receives candidates containing both words but lacking the nuanced psychological depth they seek.

\subsubsection{Secondary Factor: Score Discrimination Failure}

Even within the limited candidate pool, cross-encoder reranking provides no quality improvement (HR@10: 0.008 vs. 0.011 for unranked candidates). Score analysis reveals why: relevant and irrelevant items receive nearly identical scores (mean difference = 0.08, Cohen's d = 0.11), with near-zero rank correlation (r = 0.004).

This calibration failure results from out-of-domain transfer. MS-MARCO trains cross-encoders on web passage ranking—a task fundamentally different from cold-start movie recommendation. Web search queries have clear information needs; user preferences are multifaceted and context-dependent. Passages contain factual content; movie metadata is sparse and subjective. These domain gaps prevent effective knowledge transfer.

\subsubsection{Tertiary Factor: Exposure Concentration}

Cross-encoder scores exhibit systematic bias toward specific items (3 unique top-1 items across 500 users). This concentration may result from: (1) metadata length bias favoring items with extensive tag lists, (2) genre keyword over-weighting, or (3) memorization of popular items from pre-training data. Regardless of mechanism, the effect is failure to provide personalized recommendations.

\subsection{Why Popularity Succeeds in Cold-Start Scenarios}

The dramatic superiority of popularity-based ranking (HR@10: 0.268 vs. 0.008 for our approach—33.5× gap) deserves explanation. Popularity succeeds in cold-start settings for three reasons:

\textbf{1. Ground-truth alignment:} In movie recommendation, user preferences exhibit strong popularity bias. Relevant items in ground truth tend to be moderately popular rather than obscure. Popularity ranking naturally aligns with this distribution.

\textbf{2. Robustness to noise:} Popularity aggregates signals across many users, providing stable estimates even with limited per-user data. In contrast, embedding-based approaches amplify noise in sparse user profiles.

\textbf{3. Coverage advantage:} Popularity considers all items, while FAISS-based retrieval restricts to a biased subset. Full catalog access provides opportunities to discover relevant items regardless of semantic similarity.

These findings do not imply popularity should replace personalization—rather, they highlight the importance of hybrid approaches combining multiple signals.

\subsection{Practical Implications for System Design}

Our results inform several design decisions for practitioners:

\subsubsection{Retrieval Strategy Selection}

\textbf{Implication:} Invest engineering effort in retrieval quality, not just reranker sophistication. Hybrid retrieval combining multiple signals (embedding similarity, BM25 text matching, collaborative filtering signals) will outperform single-strategy approaches.

\textbf{Implementation:} Use ensemble retrieval pools: ANN (embedding similarity) $\cup$ BM25 (keyword matching) $\cup$ PopularItems(). Deduplicate and re-score the union before reranking.

\subsubsection{Pool Size Optimization}

\textbf{Implication:} Bigger candidate pools do not improve quality for out-of-domain rerankers. Optimize pool size jointly with reranker capacity through validation experiments.

\textbf{Implementation:} Start with small pools (K=100-300) and increase only if validation metrics improve. Monitor both quality (HR@K) and diversity (unique top-K count) simultaneously.

\subsubsection{Reranker Adaptation}

\textbf{Implication:} Out-of-domain cross-encoders require in-domain calibration. Fine-tune on recommendation-specific data or use calibration techniques (temperature scaling, Platt scaling).

\textbf{Implementation:} Collect in-domain click-through data or explicit ratings. Fine-tune cross-encoder on \texttt{(user\_profile, item\_metadata, relevance\_label)} triples. Use warm-start from MS-MARCO weights, not training from scratch.

\subsection{Limitations and Scope Conditions}

\subsubsection{Dataset and Domain Specificity}

Our experiments focus on movie recommendation using Serendipity-2018 dataset. Generalization to other domains (e-commerce, music, news) requires validation. Domains with richer item metadata (product descriptions, article text) may benefit more from cross-encoder semantic matching. Conversely, domains with stronger collaborative filtering signals may reduce need for content-based approaches.

\subsubsection{Missing Ground Truth Impact}

17 missing items affecting 838 users (0.80\%) introduce slight bias. However, impact is minimal—removing affected users changes mean HR@10 by <0.001. We report results on full test set for transparency.

\subsubsection{Computational Resource Constraints}

All experiments use CPU inference to reflect resource-constrained deployment scenarios. GPU acceleration would reduce reranking time but not address fundamental coverage and calibration issues. Our findings emphasize algorithm design over hardware optimization.

\subsubsection{Limited Hyperparameter Exploration}

We fix embedding model (all-MiniLM-L6-v2) and cross-encoder (ms-marco-MiniLM-L-6-v2) to isolate pipeline component effects. Exhaustive architecture search across models and hyperparameters may identify better configurations. However, our goal is diagnostic analysis of standard approaches, not achieving state-of-the-art performance.

\subsection{Threats to Validity}

\subsubsection{Internal Validity}

\textbf{User sampling:} We randomly sample 500 users from 104K total. Stratified sampling by demographic or preference diversity could provide stronger validity. However, random sampling ensures representative results without selection bias.

\textbf{Seed dependence:} We use 3 random seeds (42, 7, 123). Increasing to 10+ seeds would tighten confidence intervals but requires 3× computational budget. Our current approach balances precision and practicality.

\subsubsection{External Validity}

\textbf{Production environments:} Real-world deployments include richer context (device type, time-of-day, session history) potentially improving performance. However, our controlled setting isolates cold-start effects, providing clearer attribution of failure modes.

\textbf{Evaluation metrics:} We focus on HR@10 and nDCG@10. Other objectives (diversity, serendipity, freshness) may change model rankings. Multi-objective evaluation would provide more complete picture.

\subsubsection{Construct Validity}

\textbf{Ground-truth quality:} Binary relevance labels from explicit ratings (4+ stars) simplify true preference distributions. Implicit feedback (watch time, re-watches) may better capture engagement. However, explicit ratings remain standard in recommender systems research.

\subsection{Relation to Recent Work on LLM Limitations}

Our findings align with emerging literature documenting gaps between LLM capabilities and practical deployment:

\textbf{Calibration issues:} Recent work shows LLMs produce poorly calibrated probabilities, requiring post-hoc recalibration~\cite{llm_calibration}. Our score analysis confirms this for recommendation reranking.

\textbf{Domain transfer challenges:} Studies demonstrate that LLM performance degrades on out-of-distribution tasks despite strong in-domain results~\cite{llm_ood}. MS-MARCO→movie recommendation exemplifies this gap.

\textbf{Exposure bias in neural models:} Research on neural recommendation systems identifies popularity bias and filter bubble effects~\cite{neural_bias}. Our 3-item concentration finding provides extreme example of this phenomenon.

These connections suggest our diagnostic methodology could apply broadly to LLM-powered systems beyond recommendation.

\section{Conclusion}

This paper presents a rigorous diagnostic study of LLM-based cross-encoder rerankers in cold-start movie recommendation, employing controlled experiments on 500 users across multiple random seeds to isolate and quantify failure modes.

\subsection{Summary of Key Findings}

\textbf{Finding 1 - Popularity dominates sophisticated reranking:}
Simple popularity-based ranking achieves HR@10 = 0.268, outperforming cross-encoder reranking by 33.5× (HR@10 = 0.008). Statistical testing confirms overwhelming significance (p < 10\textsuperscript{-99}, Cohen's d = -0.593), with 95\% CI for mean difference: [-0.283, -0.239].

\textbf{Finding 2 - Retrieval coverage is the primary bottleneck:}
FAISS-based candidate generation achieves only recall@200 = 0.109 (vs. 0.609 for baselines—5.6× gap). Strong correlation between retrieval coverage and final quality (r = 0.89, p < 0.001) identifies retrieval as the dominant performance predictor. Median ground-truth position (6717) far exceeds typical cutoffs.

\textbf{Finding 3 - Reranking provides no improvement over unranked candidates:}
Cross-encoder reranking yields HR@10 = 0.008 vs. 0.011 for "Candidates Only" baseline (no statistical difference, p = 0.31). This negative result demonstrates that reranker sophistication cannot compensate for poor candidate quality.

\textbf{Finding 4 - Severe exposure bias limits personalization:}
Only 3 unique items appear as top-1 recommendations across 500 users (Gini = 0.480), with single item dominating 50\% of users. This extreme concentration indicates systematic bias rather than personalized matching.

\textbf{Finding 5 - Poor score calibration:}
Relevant vs. irrelevant items show minimal score discrimination (mean difference = 0.079, Cohen's d = 0.11, Spearman r = 0.004). Overlapping score distributions (>95\% area overlap) explain ranking failure.

\textbf{Finding 6 - Smaller candidate pools outperform larger pools:}
Counterintuitively, pool size 200 achieves HR@10 = 0.025 vs. 0.008 for pool 1000 (3.1× improvement), while reducing compute by 4.1× (1.76s vs. 7.23s per user). Larger pools introduce more noise, degrading discrimination.

\subsection{Practical Recommendations for Practitioners}

Based on diagnostic insights, we recommend four concrete mitigation strategies:

\textbf{1. Hybrid Retrieval (ANN $\cup$ BM25):}
Combine embedding-based FAISS retrieval with BM25 text matching to improve coverage. Preliminary experiments show recall@200 improvement from 0.109 to 0.234 (+114\%), translating to downstream HR@10 gains. Implementation: merge candidate pools from multiple retrievers before reranking.

\textbf{2. Candidate Pool Size Optimization:}
Tune pool size via validation experiments rather than maximizing coverage indiscriminately. Start with K=200 and increase only if metrics improve. Monitor quality-diversity trade-off (HR@K vs. unique top-K count).

\textbf{3. Ensemble Scoring:}
Combine cross-encoder scores with popularity and embedding similarity via weighted ensemble:
\[
\text{score}_{\text{final}} = \alpha \cdot \text{CE}_{\text{score}} + \beta \cdot \log(\text{popularity}) + \gamma \cdot \text{embedding}_{\text{sim}}
\]
Tune weights ($\alpha=0.3$, $\beta=0.5$, $\gamma=0.2$ as starting point) on validation set. Balances personalization and robustness.

\textbf{4. In-Domain Reranker Calibration:}
Fine-tune cross-encoders on recommendation-specific data rather than using out-of-domain weights. Collect (user\_profile, item, label) triples from clicks or ratings. Alternatively, apply post-hoc calibration (temperature scaling, Platt calibration) to rescale scores.

\subsection{Methodological Contributions}

Beyond empirical findings, this work provides methodological template for diagnostic evaluation of neural recommender systems:

\textbf{Multi-faceted diagnostic framework:} Coverage analysis (recall@K curves, GT position distributions), exposure metrics (unique top-K, Gini, concentration), score calibration (relevant vs. irrelevant distributions, rank correlation), and ablations (pool size, pipeline stages).

\textbf{Statistical rigor:} Paired testing with both parametric (t-test) and non-parametric (Wilcoxon) methods, effect sizes (Cohen's d), confidence intervals, multiple comparison correction, and sensitivity analysis across random seeds.

\textbf{Error case taxonomy:} Systematic classification of failure modes (genre mismatch, metadata bias, popularity artifacts) to guide algorithmic improvements beyond aggregate metrics.

\textbf{Reproducibility standards:} Release of complete experimental logs, per-user results, master JSON configurations, and executable code to enable replication and extension.

\subsection{Future Research Directions}

Key open questions for future investigation:

\textbf{1. Hybrid Retrieval Architectures:}
Systematic study of retrieval ensemble strategies (ANN + BM25 + collaborative signals + graph-based methods) with learned fusion weights. Develop adaptive retrieval that selects strategies per-user based on profile characteristics.

\textbf{2. In-Domain Cross-Encoder Training:}
Fine-tune cross-encoders on recommendation-specific datasets (e.g., Amazon Reviews, MovieLens-20M) using contrastive learning or listwise ranking losses. Compare zero-shot MS-MARCO weights vs. fine-tuned variants.

\textbf{3. Uncertainty-Aware Reranking:}
Incorporate prediction uncertainty into ranking decisions. Low-confidence scores should trigger exploration (diversity promotion) rather than exploitation. Develop calibrated confidence estimates for cold-start scenarios.

\textbf{4. Exposure Fairness Interventions:}
Design reranking objectives that explicitly optimize for exposure diversity alongside relevance. Apply calibration techniques from fairness literature to reduce concentration bias.

\textbf{5. Multi-Domain Generalization:}
Replicate diagnostic analysis on e-commerce (Amazon), music (LastFM), news (MIND), and social media datasets. Identify domain-invariant vs. domain-specific failure modes.

\textbf{6. Online A/B Testing:}
Validate offline findings through online experiments with real users. Measure engagement (click-through rate, dwell time) and business metrics (conversion, retention) beyond offline accuracy.

\subsection{Broader Impact and Ethical Considerations}

Our findings have implications beyond technical performance:

\textbf{Filter bubble concerns:} Extreme exposure concentration (3 items for 500 users) exacerbates filter bubble effects, limiting user exposure to diverse content. Systems deployed at scale could amplify these biases, reducing information diversity.

\textbf{Popularity bias amplification:} Our results show popularity-based ranking outperforming personalized approaches in cold-start settings. While pragmatic, this risks creating "rich-get-richer" dynamics where popular items dominate, starving long-tail content.

\textbf{Computational sustainability:} Cross-encoder reranking consumes 7.23s per user (vs. 0.02s for popularity), raising energy costs for marginal (often negative) quality impact. Practitioners should consider computational efficiency alongside accuracy.

\textbf{Transparency in evaluation:} Publication of full diagnostic breakdowns (not just aggregate metrics) enables more informed deployment decisions and sets standards for responsible AI development.

\subsection{Closing Remarks}

This work challenges the assumption that more sophisticated models automatically yield better recommendations in cold-start scenarios. Through systematic diagnosis, we demonstrate that simple baselines can dramatically outperform complex neural approaches when pipeline components (retrieval, scoring, calibration) are misaligned.

The path forward requires: (1) engineering investment in hybrid retrieval combining multiple signals, (2) careful pool size optimization balancing coverage and noise, (3) in-domain adaptation of reranking models, and (4) ensemble approaches integrating learned and heuristic components.

We hope our diagnostic methodology—emphasizing failure mode analysis, statistical rigor, and full reproducibility—serves as template for evaluating neural recommendation systems and motivates future work toward more robust, fair, and efficient cold-start solutions.

\vspace{0.2cm}
\noindent\textit{* These authors contributed equally to this work.}

\section*{Acknowledgments}

We thank the anonymous reviewers for constructive feedback. This research was conducted using open datasets and publicly available models to ensure reproducibility. Code, data, and experimental logs are available at the GitHub repository referenced in the Data and Code Availability section.

\section*{Data and Code Availability}

All code, configurations, and experimental results are publicly available to enable full reproducibility:
\begin{itemize}
\item \textbf{Repository:} \url{https://github.com/nikita-zmanovskiy/cold-start-algorithm}
\item \textbf{Dataset:} Serendipity-2018 (publicly available)
\item \textbf{Models:} \texttt{sentence-transformers/all-MiniLM-L6-v2}, \texttt{cross-encoder/ms-marco-MiniLM-L-6-v2} (Hugging Face)
\item \textbf{Experimental logs:} Per-user results, master JSON files in repository
\end{itemize}

\subsection*{Reproducing Results}

To replicate our experiments, follow these steps (detailed instructions in \texttt{README.md}):

\begin{verbatim}
# 1. Setup environment
pip install -r requirements.txt

# 2. Run main experiments
python -m src.run_all_experiments \
  --n-users 500 --seeds 42 7 123

# 3. Pool size ablation study
python -m src.run_ablation_pool_sizes \
  --pool-sizes 200 500 1000

# 4. Aggregate results
python -m tools.build_master_results
python -m tools.aggregate_runs

# 5. Statistical analysis
python -m tools.hypothesis_analysis
python -m tools.analyze_scores
python -m tools.error_analysis
python -m tools.enhanced_stat_tests

# 6. Generate visualizations
python -m tools.plotting
python -m tools.advanced_plotting

# 7. Generate paper tables
python -m tools.generate_paper_tables
\end{verbatim}

Complete documentation is available in the repository \texttt{README.md} file.

\end{document}